\documentclass[prl,twocolumn,showpacs]{revtex4}
\usepackage{epsfig,multirow}

\begin{document}
\title{Spin rotations induced by electron running on closed
trajectories in gated semiconductor nanodevices}
\author{S. Bednarek} \affiliation{Faculty of Physics and Applied
Computer Science, AGH University of Science and Technology,\\ al.
Mickiewicza 30, 30-059 Krak\'ow, Poland}
\author{B. Szafran}
\affiliation{Faculty of Physics and Applied Computer Science, AGH
University of Science and Technology,\\ al. Mickiewicza 30, 30-059
Krak\'ow, Poland}

\begin{abstract}
A design for a quantum gate performing transformations of a single
electron spin is presented. The spin rotations are performed by the
electron going around the closed loops in a gated semiconductor
device. We demonstrate the operation of NOT, phase-flip and Hadamard
quantum gates, i.e. the single-qubit gates which are most commonly
used in the algorithms. The proposed devices employ the
self-focusing effect for the electron wave packet interacting with
the electron gas on the electrodes and the Rashba spin-orbit
coupling. Due to the self-focusing effect the electron moves in a
compact wave packet. The spin-orbit coupling translates the spatial
motion of the electron into the rotations of the spin. The device
does not require microwave radiation and operates using low constant
voltages. It is therefore suitable for selective single-spin
rotations in larger registers.
\end{abstract}
 \maketitle Extensive work for
design and construction of quantum processing devices is underway.
The potential implementations are based on various effects and
systems including photonic \cite{ph} and superconducting \cite{sp}
devices, nuclear magnetic resonance \cite{nmr} and ion traps
\cite{it}. In one of the approaches the quantum bits are stored by
spins of electrons confined in semiconductor nanostructures
\cite{sp1}. Such a quantum gate can be naturally combined with a
classical computer.
So far the set-up and read-out \cite{read} of the spin were realized
as well as the spin rotations \cite{rotate,rot4,rot5} in single and
double dots.

According to the original proposal \cite{sp1} a universal quantum
gate requires exchange operations between pairs of spins combined
with the single-spin rotations. The latter can be performed by the
Rabi oscillations in an external microwave field. For a number of
reasons selective single-spin rotations are considered more
challenging to implement than the spin exchange between a pair of
electrons \cite{nature}. The problem with the microwave radiation is
that even at magnetic fields of the order of 10 T the spin Zeeman
energy splitting is relatively low and the resonant wavelength is of
the order of millimeters, which rather excludes fast and
site-selective operations on a single spin without perturbing the
others. It was therefore suggested that a universal two-qubit gate
can be achieved applying the Heisenberg coupling only:  employing
additional registers \cite{nature}, using inhomogenous Zeeman
splitting \cite{levy} or exploiting the spin-orbit coupling
\cite{stepanenko}. Experimentally a site-selective single-spin
rotations were eventually demonstrated with an embedded local
microwave source \cite{rot4} which however requires cooling of the
heat generated by the AC currents. The cooling problem is avoided
when the spin rotations are induced by oscillating electric fields
\cite{rot5} and occur due to the spin-orbit coupling.  In this
Letter we propose a device in which the single-spin operations are
performed without the microwave radiation or fast voltage
oscillations. The proposed device is based on spatial motion of the
confined electron in the presence of the spin-orbit coupling and
requires application of low DC voltages only.

\begin{figure}[ht!]
\centerline{\hbox{\epsfysize=30mm
               \epsfbox[98 590 260 700] {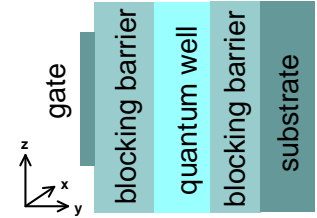}\hfill}}
\caption{(color online) Schematics of the consider semiconductor
structure.
 \label{ss}}
\end{figure}

We recently showed that induced quantum wires and dots \cite{a1} are
formed under metal gates deposited on a planar structure containing
a quantum well due to  the self-focusing effect \cite{a2,a3} for the
wave function of the confined electron interacting with the electron
gas in the metal. This effect assists in the 100\% guaranteed
transfer of a stable electron packet following a trajectory which is
controlled by the gate set-up and applied DC voltages.

We consider a planar nanostructure of Fig. 1 with a quantum well and
electrodes on top. A single electron is confined in the quantum
well. We assume that the quantum well is made of a semiconductor of
the diamond lattice structure (Si, Ge), in which the Dresselhaus
spin-orbit coupling is absent  due to the inversion symmetry of the
crystal. The electron motion in the $y$ direction is frozen by the
quantum well confinement. We use a two-dimensional Hamiltonian
\begin{equation}
H(x,z,t) =-\frac{\hbar^2}{2m}\left(\frac{\partial ^2}{\partial
x^2}+\frac{\partial ^2}{\partial z^2}\right)-e\phi_2(x,y_0,z,t)+H_R,
\end{equation}
where $y_0$ is the coordinate of the center of the quantum well and
$H_R$ is the Rashba
spin-orbit coupling term due to the
asymmetry of the quantum well potential
$H_R=\alpha\left(p_z\sigma_x-p_x\sigma_z\right),
$
where $p$ are the momentum operators and $\sigma$'s are the Pauli
matrices. We write the state functions as vectors (spinors)
\begin{equation}
\Psi(x,z,t)=\left(\begin{array}{c} \psi_1(x,z,t) \\
\psi_2(x,z,t) \end{array} \right).
\end{equation}

\begin{figure}[ht!]
 \centerline{\hbox{\epsfysize=55mm
               \epsfbox[97 545 366 764] {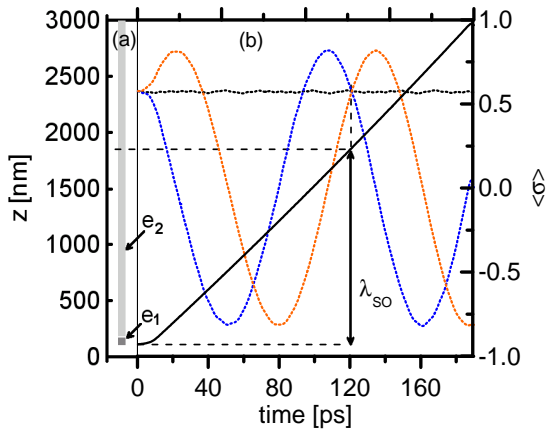}\hfill}}
\caption{(color online) (a) System of two electrodes ($e_1,e_2$) on
top of the structure. (b) Electron packet $z$ position vs time (dark
solid line, left axis). Dotted lines show average values of the
Pauli operators: $\langle\sigma_z\rangle$ (red color),
$\langle\sigma_x\rangle$  (black) and $\langle\sigma_y\rangle$
(blue) and are referred to the right axis. \label{ssd}}
\end{figure}


The electrostatic potential $\phi_2$ of Eq. (1) is found from the
Poisson equation using the methodology previously applied for
simulations of electrostatic quantum dots \cite{e1}. $\phi_2$ is the
difference of the total electrostatic potential $\Phi$ and the
electron self-interaction potential $\phi_1$,
$\phi_2({\bf r},t)=\Phi({\bf r},t)-\phi_1({\bf r},t).$
The total potential fulfills  the 3D Poisson equation
\begin{equation}
\nabla^2\Phi({\bf r},t)=-{\rho({\bf r},t)}/{\epsilon\epsilon_0},
\end{equation}
and the self-interaction potential is calculated with the Coulomb
law
\begin{equation}
\phi_1({\bf r},t)=\frac{1}{4\pi\epsilon\epsilon_0} \int d{\bf r}'
\frac{\rho({\bf r}',t)}{|{\bf r}-{\bf r}'|},
\end{equation}
where $\rho({\bf r},t)$ is the electron density calculated for wave
function (2)
\begin{equation}
\rho({\bf r},t)=-e\left(|\psi_1(x,z,t)|^2 +
|\psi_2(x,z,t)|^2\right)\delta(y-y_0).
\end{equation}
Eq. (3) is solved numerically in a rectangular box containing the
studied nanostructure. Potentials applied to the gates are assumed
as Dirichlet boundary condition.  The content of the computational
box is charge neutral, so on its surface we assume vanishing normal
component of the electric field. Calculated potential $\phi_2({\bf
r},t)$ contains a contribution of the charge induced on the metal
surface by the confined electron. This contribution introduces the
self-focusing effect \cite{a2}. The time dependence in (3-5) enters
due to the motion of the electron packet. As the initial condition
we take the solution of a time-independent Schr\"odinger equation
$H(x,z)\Psi_0(x,z)=E\Psi_0(x,z)$ for a given spin-state. The time
evolution
is obtained numerically with a finite-difference scheme consistent
with the time-dependent Schr\"odinger equation
$
\Psi(t+dt)=\Psi(t-dt)-\frac{2i}{\hbar} H(t)\Psi(t)dt.
$

\begin{figure}[ht!]
 \centerline{\hbox{\epsfysize=45mm
               \epsfbox[80 550 344 740] {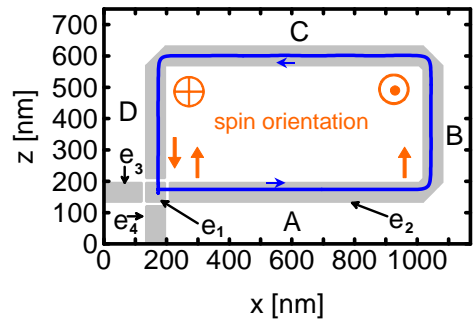}\hfill}}
\caption{(color online) Gate configuration (grey colors) for the NOT
gate. Blue solid lines show the electron trajectory (electron starts
from under $e_1$ and goes to the right). Red symbols show the spin
orientation near the corners of the trajectory ($\bigodot$,
$\bigoplus$ indicate ``from the page'' and ``to the page''
directions, respectively). \label{ssx}}
\end{figure}

\begin{figure}[ht!]
 \centerline{\hbox{\epsfysize=50mm
               \epsfbox[92 538 387 781] {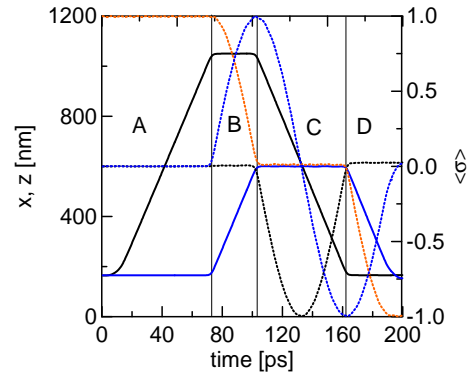}\hfill}}
\caption{(color online) Same as Fig. 2 but for the NOT gate of Fig.
3.                The black (blue) solid curves show the $x$, ($z$)
               positions as a function of time.
\label{sswd}}
\end{figure}

Applying weak voltages to the gates with respect to the substrate
one can \cite{a1} set the electron packet in motion and stop it in a
chosen location. The products of the momentum and spin operators in
the $H_R$ operator perturb somewhat the electron trajectories.
Electron motion influences the spin in a much more pronounced
extent.

Let us consider the system presented in Fig. 2(a) with two
electrodes $e_1$ and $e_2$ placed parallel to the $z$ axis on top of
the structure of Fig. 1. In the initial state we put zero voltage to
electrode $e_1$ and small negative to $e_2$ \cite{a1}. The
ground-state wave function is formed under $e_1$. We assume that the
spin is in the state which has the same average value in all $x,y,z$
directions,
$\Psi(x,z,0)=\Psi_0(x,z)\frac{1}{\sqrt{2}\sqrt{\sqrt{3}+3}}\left[\begin{array}{c}(1+\sqrt{3})\\
(1+i)\end{array}\right]. $
The motion of the packet starts when the voltage applied to $e_2$ is
switched to $V_2=0.2$ mV which extracts the electron from underneath
the gate $e_1$. The electron is initially accelerated when it passes
from under $e_1$ to under $e_2$, then it moves with a constant
velocity. The time dependence of the electron position is given in
Fig. 2(b) with a solid black line. The dotted curves show the
average values of the components of the spin. We see that for the
 electron moving parallel to the $z$ axis the $\langle\sigma_x\rangle$ value is
 preserved and the  $\langle\sigma_y\rangle$ and  $\langle\sigma_z\rangle$ components
 oscillate: the spin is rotated around the $x$ axis.
 Similarly,  for the electron moving along the $z$
 direction the Rashba coupling leads to the spin rotation around
 the $x$ axis.
 The rotation angle depends on
 the coupling constants, the electron effective mass, and the distance traveled by the electron.
 For simulations presented in Fig. 3 we assume the coupling constant
 $\alpha=7.2 \times 10 ^{-13}$ eVm within the range predicted for
 the asymmetric quantum wells \cite{eS}. The quantum well
 and the potential barriers are taken 10 nm thick. We apply the Si material parameters $m=0.19m_0$ and
 $\epsilon=13$. We deduce that the distance for which the initial
 spin is restored is $\lambda_{SO}=1.8$ $\mu$m.

\begin{figure}[ht!]
 \centerline{\hbox{\epsfysize=60mm
               \epsfbox[74 464 323 776] {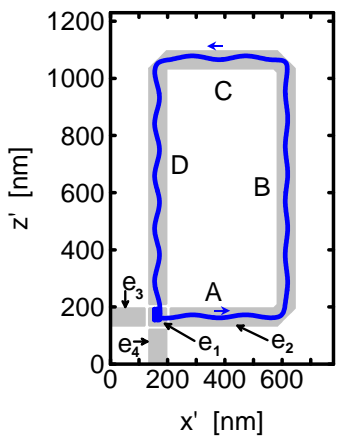}\hfill}}
\caption{(color online) Gate configuration (grey colors) for the
Hadamard gate. Blue solid lines show the electron trajectory
(electron starts from under $e_1$ and goes to the right -- the
direction of motion is marked with the blue arrows). The rectangular
gate path is rotated by 45$^\circ$ with respect to Fig. 3:
$z'=\frac{1}{\sqrt{2}}\left(x+z\right),
x'=\frac{1}{\sqrt{2}}\left(x-z\right).$
 \label{sdxsx}}
\end{figure}

\begin{figure}[ht!]
 \centerline{\hbox{\epsfysize=50mm
               \epsfbox[92 538 387 781] {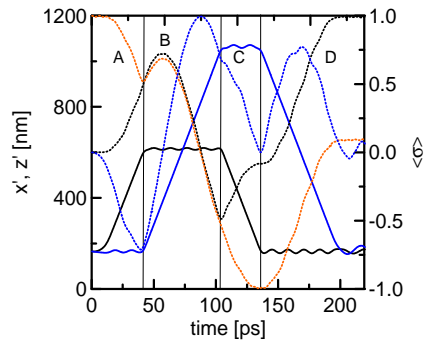}\hfill}}
               \caption{(color online) Position of the electron
               packet as a function of time for the Hadamard gate of
               Fig. 5 in the rotated system of coordinates ($x',y',z'$).
               The black (blue) solid curves show the $x'$, ($z'$)
               positions.
               The dotted curve show the expectation value
               of the Pauli matrix operators defined with respect to the
               $x,y,z$ axes.
 \label{sdxsxx}}
\end{figure}

Since the electron motion along perpendicular directions induces
spin rotations around perpendicular axes, one can perform any
rotation by making the electron move under electrodes forming a
closed loop. In Fig. 3 we propose a setup performing the logical NOT
operation on the electron spin. The electrodes are marked with the
grey color. The spin of the electron confined in the quantum dot
induced \cite{a1} under $e_1$ electrode stores the qubit. The $e_2$
electrode serves to guide the electron around a closed loop back to
the dot induced under $e_1$. For illustration initially the
z-component of the spin is set in the ``up'' state
$\Psi(x,z,0)=\left(\begin{array}{c} \Psi_0(x,z)\\ 0
\end{array} \right).$
The packet is set in motion to the right by applying a constant +0.2
mV voltage to $e_2$ and a short pulse of $-0.4$ mV to $e_3$. The
electron trajectory is drawn with the blue curve in Fig. 3.
The time dependence of the electron position is plotted in Fig. 4
(black curve shows the $x$ position and the blue one the $z$
position). In the A region we notice an initial increase of the
velocity and then a constant velocity motion till the end of the A
segment. After reflection at the cut corner \cite{a1} the electron
goes into the $B$ part where the $x$ position becomes fixed and the
$z$ one increases with time. Passing under the C segment the
electron returns to its initial $x$ coordinate and under the D
segment to its initial $z$ coordinate. At the end of the electron
slows down which results of the $e_1$, $e_2$ potential difference.
When the electron comes to under $e_1$ the potential of this
electrode is changed to $+0.3$ meV which traps the electron in the
induced dot. The oscillations of the blue curve at the end of the
motion are due to an excess of the kinetic energy.

The spin direction at the corners of the loop is schematically
marked by arrows in Fig. 3. The time dependence of
$\langle\sigma_x\rangle$, $\langle\sigma_y\rangle$ and
$\langle\sigma_z\rangle$ are plotted in Fig. 4 with dotted lines:
black, blue and red, respectively.  Initially the spin is oriented
``up'' $\langle\sigma_z\rangle=1$, and
$\langle\sigma_x\rangle=\langle\sigma_y\rangle=0$.  In the A segment
the electron moves in the $x$ direction so the spin is rotated
around the $z$ axis and no spin change is observed in Fig. 4. The
length of the $B$ segment is such that the spin is rotated around
the $z$ axis by $90^\circ$ and takes the ``from the page''
orientation: the spin is in the $\sigma_y$ eigenstate and
$\langle\sigma_y\rangle=1$. When the electron returns in the $-x$
direction the spin is rotated by $180^\circ$ degrees and takes the
"to the page" orientation $\langle\sigma_y\rangle=-1$  at the end of
the $C$ part. On the $D$ segment the spin is rotated by $-90^\circ$
degrees around the $x$ axis. Returning to $e_1$ the electron is in
the  ``down'' spin eigenstate $\langle\sigma_z\rangle=-1$. Similarly
one can show that the same trajectory inverts the spin of initial
``down'' orientation.  Thus the motion around the loop performs the
NOT operation
\begin{equation}
U^{\mathrm{NOT}}=\left(\begin{array}{cc} 0 & 1  \\ 1 & 0 \end{array}
\right).
\end{equation}

In the theory of quantum computation other useful single qubit
operations are the Hadamard transformation $U^H$ of the basis states
into their equilibrated superpositions and the phase flip operation
$U^\pi$
\begin{equation}
U^H=\frac{1}{\sqrt{2}} \left(\begin{array}{cc} 1 & 1  \\ 1 & -1
\end{array} \right), U^\pi=\frac{1}{\sqrt{2}} \left(\begin{array}{cc} 1 & 0  \\ 0 & -1
\end{array} \right).
\end{equation}
The loop of Fig. 3 can be used as the Hadamard gate for the spin
``up'' and ``down'' states redefined with respect to the direction
bisecting the angle between $x$ and $-z$ axes. Alternatively one can
keep the basis set and rotate the electron trajectory (the electrode
loop) by 45 degrees in the $(x,z)$ plane. The electrode system
corresponding to this gate and the electron trajectory are depicted
in Fig. 5. The plot is draw in rotated coordinate system
$z'=\frac{1}{\sqrt{2}}\left(x+z\right)$,
$x'=\frac{1}{\sqrt{2}}\left(x-z\right)$. The simulation of the
Hadamard gate operation was performed similarly as the one
performing the NOT operation. As the initial condition we took the
electron confined below the $e_1$ electrode in the ground-state with
the spin parallel to the $z$ axis. The packet is set in motion to
the right by introducing a potential difference between $e_1$ and
$e_2$ equal to $-0.2$ mV. This time one cannot illustrate the spin
orientation near the corners of the trajectory because the rotation
angles are not multiples of $90^\circ$ and in two first corners the
spin projections on the $x,y,z$ axes are not definite. This can be
clearly seen in Fig. 6 which shows the time dependence of the spin
average values. The spin initially oriented ``up''
$\langle\sigma_z\rangle=1$, after closing the trajectory loop is set
to the ``right'' $\langle\sigma_x\rangle=1$. Similarly, the spin
``down'' $\langle\sigma_z\rangle=-1$ turns to the ``left'' at the
end of the loop. Twofold rotation around the loop is equivalent to
the identity transform, i.e. the rotation by the full angle.

The phase-shift operation $U^\pi$ is performed by the electrode
configuration rotated by a $90^\circ$ angle with respect to the NOT
gate of Fig. 3. The NOT quantum gate oriented as in Fig. 3 performs
the $U^\pi$ transformation for the ``up'' and ``down'' states
redefined as parallel and antiparallel to the $x$ axis.

The devices proposed here use the Rashba coupling in the absence of
the Dresselhaus interaction. These devices can be realized in
semiconductors of the diamond lattice (Si, Ge). The Rashba coupling
can be tuned by external electric field and it becomes arbitrarily
small for nearly symmetric quantum wells. For devices based on zinc
blende materials  (III-V's or II-VI's) one cannot get rid of the
Dresselhaus term but one can still design electron trajectories
performing the spin operations for both couplings present
\cite{tbp}.

We demonstrated that the controlled electron motion around closed
loops along induced quantum wires combined with the spin-orbit
coupling can be used to design devices performing any single-spin
rotation. The proposed device is scalable and since it runs without
a microwave radiation or high frequency electric fields it offers an
independent control of many separate qubits.

\end{document}